\documentclass[journal=jpclcd,manuscript=letter]{achemso}
\usepackage{amsmath,amssymb}
\usepackage{float}

\usepackage[T1]{fontenc}
\usepackage[utf8]{luainputenc}
\usepackage{float}
\usepackage{multirow}
\usepackage{lipsum}
\usepackage{mathrsfs}
\usepackage{setspace}
\usepackage{droidsans}
\usepackage{balance}
\usepackage{times,mathptmx}
\usepackage{braket}
\usepackage{sectsty}
\usepackage{graphicx}
\usepackage{amsmath,amssymb}
\usepackage{mathrsfs}
\usepackage{multirow} 
\usepackage{lastpage}
\usepackage{adjustbox}
\usepackage{upgreek}
\usepackage{color}
\usepackage{array}
\usepackage{float}
\usepackage{fancyhdr}
\usepackage{fnpos}
\usepackage{adjustbox}
\usepackage[english]{babel}
\DeclareUnicodeCharacter{0308}{~}
\usepackage[version=3]{mhchem} 

\author{Ankita Phutela}
\email{Ankita@physics.iitd.ac.in[AP]}
\author{Preeti Bhumla, Manjari Jain}
\author{Saswata Bhattacharya}
\email{saswata@physics.iitd.ac.in[SB]}
\phone{+91-11-2659 1359}
\affiliation[Indian Institute of Technology Delhi]
{Department of Physics, Indian Institute of Technology Delhi, New Delhi, India}
\title[An \textsf{achemso} demo]
{Exploring strong and weak topological states on isostructural substitutions in TlBiSe$_2$ }
\keywords{Topological Insulator, Surface States, Z$_2$ invariant, Spin-Orbit Coupling, DFT}
\begin{document}
	\sloppy

\begin{abstract}
	
Topological Insulators (TIs) are unique materials where insulating bulk hosts linearly dispersing surface states protected by the Time-Reversal Symmetry (TRS). These states lead to dissipationless current flow, which makes this class of materials highly promising for spintronic applications. Here, we predict new TIs via high-throughput screening by employing state-of-the-art first-principles based methodologies, viz., Density Functional Theory (DFT) and many-body perturbation theory (G$_0$W$_0$) combined with Spin-Orbit Coupling (SOC). For this, we take a well-known 3D TI, TlBiSe$_2$ and perform complete substitution with suitable materials at different sites to check if the obtained isostructural materials exhibit topological properties. Subsequently, we scan these materials based on SOC-induced parity inversion at Time-Reversal Invariant Momenta (TRIM). Later, to confirm the topological nature of selected materials, we plot their surface states along with calculation of Z$_2$ invariants. Our results show that GaBiSe$_2$ is a Strong Topological Insulator (STI). Besides, we report six Weak Topological Insulators (WTIs) viz. PbBiSe$_2$, SnBiSe$_2$, SbBiSe$_2$, Bi$_2$Se$_2$, TlSnSe$_2$ and PbSbSe$_2$. We have further verified that all the reported TIs are dynamically stable showing all real phonon modes of vibration. 
	\begin{tocentry}
		\begin{figure}[H]%
			\includegraphics[width=0.9\columnwidth,clip]{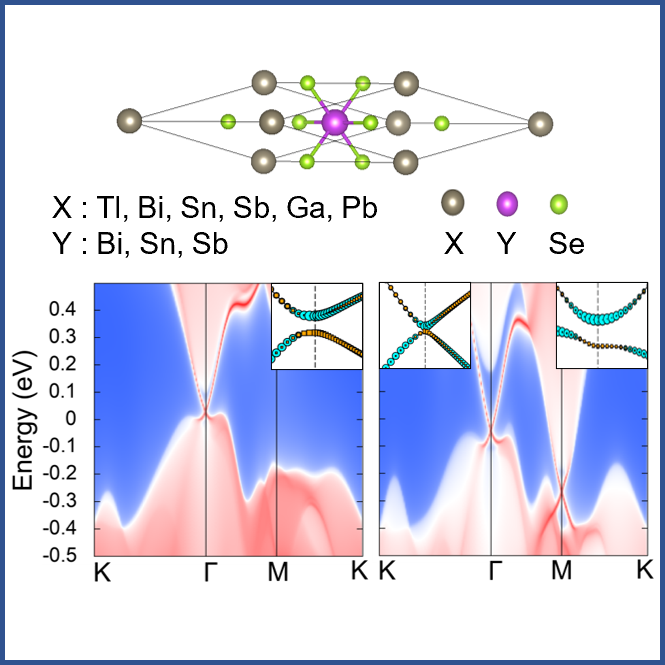}
		\end{figure}	
	\end{tocentry}
\end{abstract}

	
\section{Introduction}
Since the discovery of Topological Insulators (TIs) about a decade ago, there has been an enormous increase in interest towards topological condensed matter systems\cite{qi2008topological,moore2007topological,hasan2010colloquium,moore2010birth,fu2007topological,bernevig2006quantum,qi2011topological,hsieh2009tunable}. TIs show great potential applications in quantum computing and spintronics due to the insensitivity of the transport property towards non-magnetic perturbations\cite{wan2011topological,qi2011topological}. TIs also pave the way for realizing novel quantum phenomena such as Weyl semimetals\cite{singh2012topological}, Majorana-fermions\cite{wilczek2009majorana} and Higgs mechanism\cite{sato2011unexpected}. These alluring materials are insulating in bulk but support the flow of electrons on their surface. As a result, their surface consists of linear states that are protected by the Time-Reversal Symmetry (TRS)\cite{kane2005z}. A necessary condition for the appearance of these states is the inversion of bands, which takes place at the Time-Reversal Invariant Momenta (TRIM) in the bulk Brillouin Zone (BZ)\cite{eremeev2010ternary,singh2014topological}. The natural ordering of the energy levels forming the edges of the gap is inverted owing to the strong Spin-Orbit Coupling (SOC) associated with heavy elements.
The topology of TRS invariant insulators is characterized by the Z$_2$ index, $\nu_0$, which can either be 0 or 1, depicting a topologically trivial or non-trivial phase, respectively. However, it has been recently reported that even when $\nu_0$ = 0, the system can show non-trivial characteristics~\cite{lee2021discovery}. Based on whether the material hosts odd or even number of Dirac cones in the electronic structure of its surface, TIs are further classified as Strong Topological Insulator (STI) having Z$_2$ invariant, $\nu_0$ = 1 or Weak Topological Insulator (WTI) with Z$_2$ invariant, $\nu_0$ = 0\cite{rusinov2016mirror,majhi2017emergence}. Nevertheless, the complete characterization of 3D TIs requires a set of, in total, four Z$_2$ numbers: ($\nu_0$;$\nu_1$$\nu_2$$\nu_3$). The indices $\nu_1$, $\nu_2$ and $\nu_3$ are called weak indices and are believed to be nonrobust quantities\cite{imura2011weak,noguchi2019weak}. Therefore, the calculation of surface states and Z$_2$ invariant give complete information about the topological nature of material.

First-principles calculations led to the prediction of a large number of 2D and 3D TIs\cite{das2016pedagogic,hsieh2012topological,kou2017two}. Among the various well established families of 3D TIs, Bi$_2$Se$_3$ and Bi$_2$Te$_3$ have been most widely studied for investigating topological states and their properties\cite{zhang2009topological}. Their crystal structures consist of quintuple layers held together by weak van der Waals (vdW) forces, providing natural cleavage planes without breaking strong bonds. The band structure calculations in Bi$_2$Se$_3$ have shown that the Dirac point of surface state lies close to the valence band maximum (VBM)\cite{reid2020first}. This leads to the opening of electron scattering channel from surface states to bulk continuum states, and the topological transport regime begins to collapse. Therefore, there is a strong need for new materials with ideal and relatively isolated Dirac cones. A variety of candidates with non-trivial electronic states including HgTe\cite{brune2011quantum}, InAs\cite{lin2013adiabatic}, ternary tetramytes
Ge$_m$Bi$_{2n}$Te$_{(m+3n)}$\cite{linnature}, half-Huesler compounds\cite{lin2010half,chadov2010tunable,feng2010half,xiao2010half,shitade2009quantum,chadov2010j}, LiAuSe honeycomb lattice\cite{zhang2011topological}, $\beta$-Ag$_2$Te\cite{lee2012single} to non-centrosymmetric BiTeX (X=Cl, Br, I)\cite{chen2013discovery,landolt2013bulk,bahramy2012emergence} have been suggested. Theoretical studies have shown that Tl-based ternary chalcogenides viz., TlSbTe$_2$, TlBiSe$_2$ and TlBiTe$_2$ are 3D TIs with a single Dirac cone surface state at the $\Gamma$ point, which is well isolated from the bulk continuum\cite{lin2010single,yan2010theoretical}. Tl-based materials have a 3D character because each Tl (Bi) layer is sandwiched between two Se layers with strong coupling between neighboring atomic layers instead of weak vdW forces as in Bi$_2$Se$_3$. The electronic structure of many Tl-based TIs, viz. TlAB$_2$ (A = Sb, Bi and B = Se, Te, S), have been investigated by Density Functional Theory (DFT) calculations\cite{singh2014topological,zhang2015emergence}. The role of surface termination has also been explored in TlBiSe$_2$ and TlBiTe$_2$\cite{singh2016role}. Following this the isostructural substitution of the above base material is been endeavoured to retain it's topological properties. For example, In-based compounds like InBiTe$_2$ and InSbTe$_2$, crystallizing in the TlBiSe$_2$ like crystal structure have been studied. Surprisingly, these materials lack the Dirac cone feature, which depicts their topologically trivial nature\cite{eremeev2011ab}. Therefore, despite TlAB$_2$ (A = Sb, Bi and B = Se, Te, S) shows topologically non-trivial band structure, any other isostructural substitution to retain its topological nature is hitherto unknown.  

Motivated with this idea, in this letter, we have explored the possibility of having new materials belonging to same class of ternary chalcogenides via a thorough isostructural substitutions. First, we have performed substitution at suitable sites of TlBiSe$_2$ and scanned for those materials whose band structure shows band inversion at odd/even number of TRIMs. After that, to determine the accurate band gap of TlBiSe$_2$, we have employed various exchange-correlation ($\epsilon_{xc}$) functionals viz., PBE+SOC, HSE06+SOC, G$_0$W$_0$@PBE+SOC and G$_0$W$_0$@HSE06+SOC. The band gap obtained from G$_0$W$_0$@PBE+SOC functional is in close agreement with the experimental value. Therefore, we have further calculated the band gap of all materials using G$_0$W$_0$@PBE+SOC. Subsequently, we have examined the potential materials for their dynamical stability. The stable materials are then characterised as STI/WTI depending on whether they show odd/even number of surface states, respectively. To confirm their topological nature, we have also calculated the Z$_2$ topological invariants.

\begin{figure*}[htp]
	\includegraphics[width=0.8\textwidth]{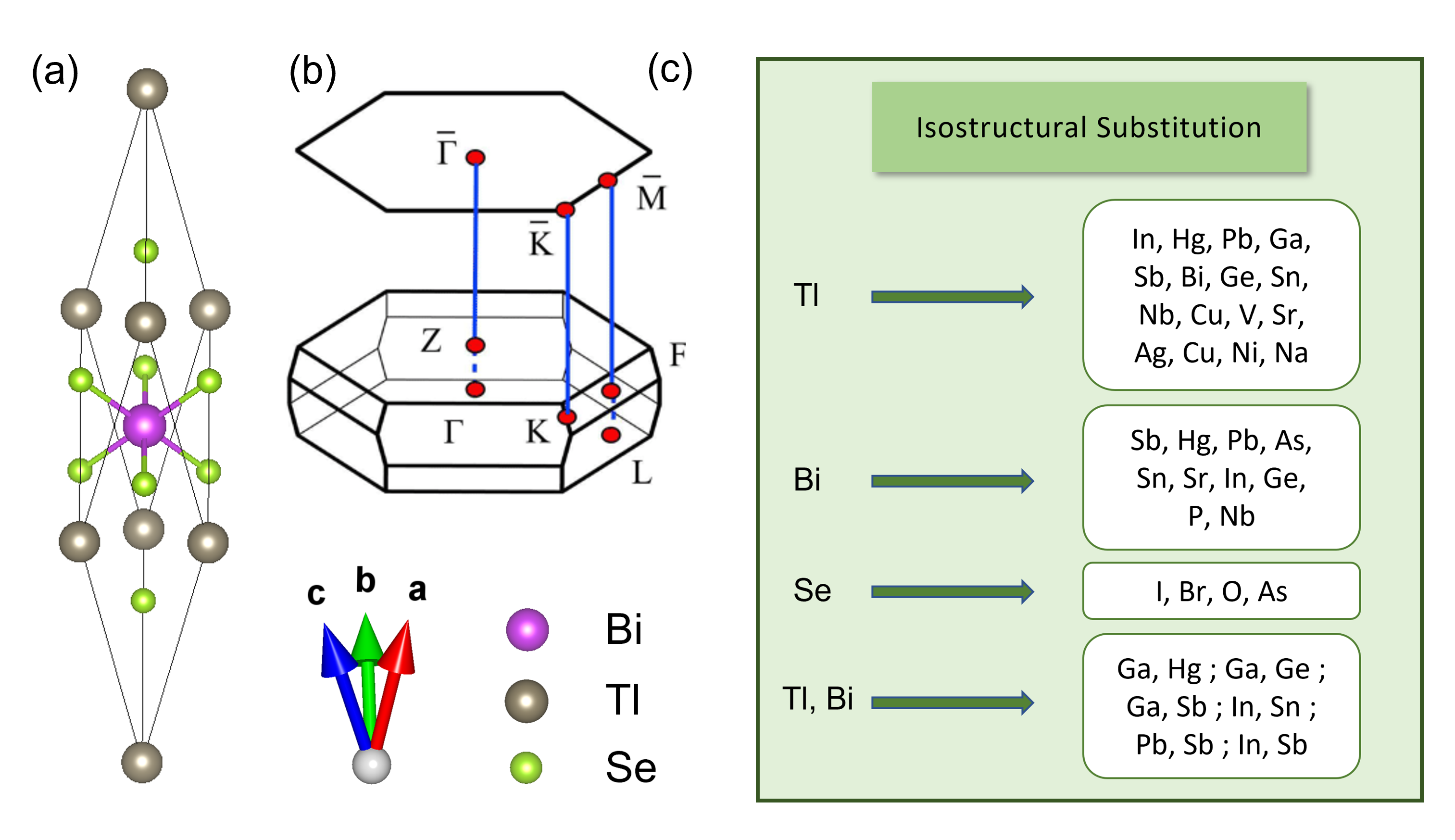}
	\centering
	\caption{(a) Primitive crystal structure of TlBiSe$_2$, (b) 3D BZ for primitive unit cell with four time-reversal invariant points $\Gamma$, Z, F and L along with the projected surface BZ, and (c) Complete substitution with various elements at Tl, Bi, Se and Tl, Bi sites simultaneously.}
	\label{pic1}
\end{figure*}

\begin{figure*}[htp]
	\includegraphics[width= 1\textwidth]{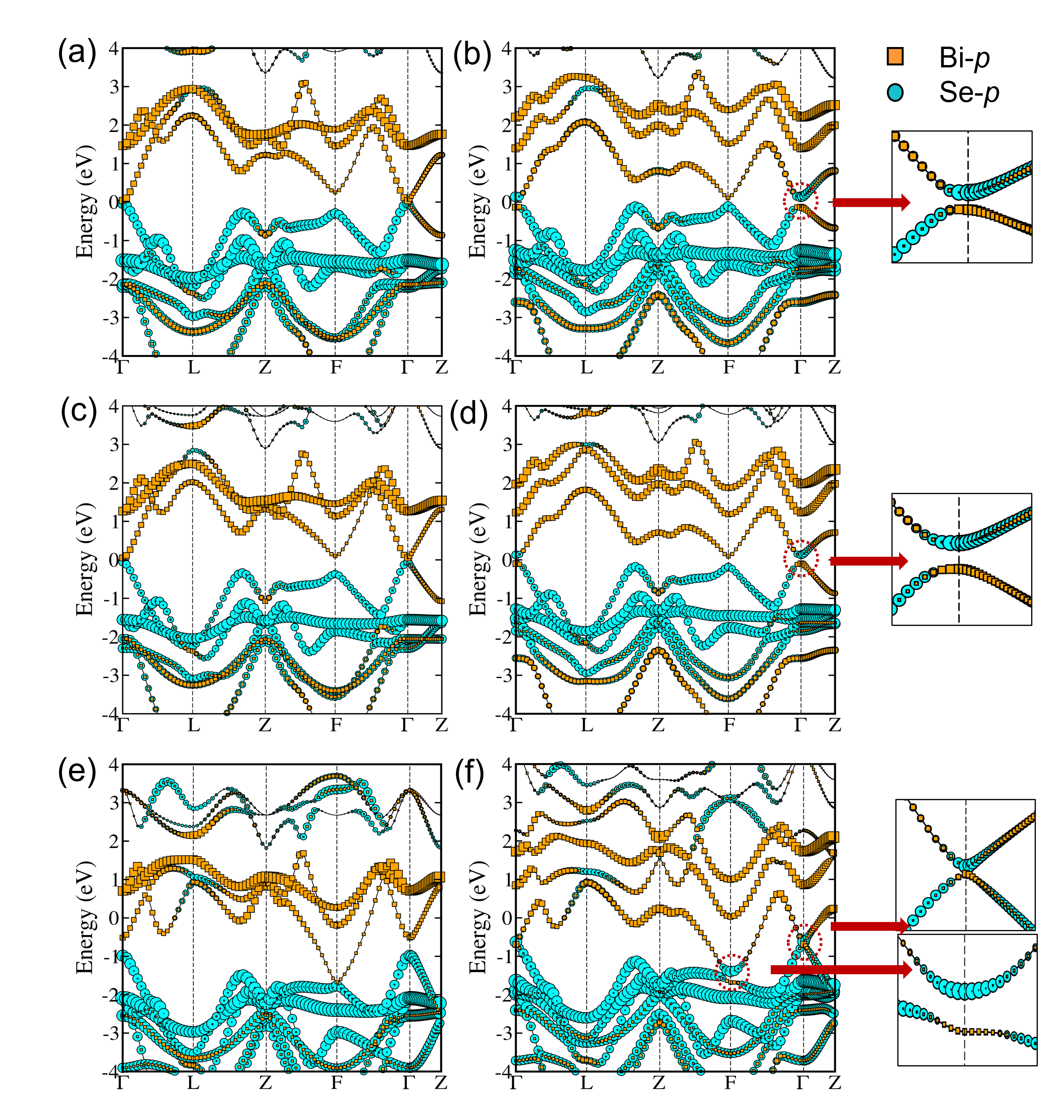}
	\centering
	\caption{The band structures for TlBiSe$_2$, GaBiSe$_2$ and PbBiSe$_2$ without SOC are shown in (a), (c), (e) and with SOC are shown in (b), (d), (f), respectively. Insets show band inversion at respective \textit{k}-points.}
	\label{pic2}
\end{figure*}

TlBiSe$_2$ belongs to the Tl-family of compounds having a rhombohedral crystal structure with space group \textit{R}$\bar{3}$\textit{m}\cite{yan2010theoretical,lin2010single}. There are four atoms in the primitive unit cell which are placed in layers normal to the three-fold axis with the sequence -Tl-Se-Bi-Se-, i.e., along [111] axis of rhombohedral unit cell (see Figure \ref{pic1}(a)). The 3D BZ for rhombohedral unit cell having high symmetry points F, $\Gamma$, L and Z, along with its projected (111) surface BZ is shown in Figure \ref{pic1}(b). The structure has inversion symmetry where both Bi and Tl act as inversion centers. We have first estimated the band gap of TlBiSe$_2$ using PBE+SOC. The calculated band gap is 215 meV (Direct), whereas, the experimental band gap is 350 meV\cite{sato2010direct}. The band gap is thus underestimated due to the DFT limitation arising from the approximations used in the $\epsilon_{xc}$ functional. Therefore, we have used hybrid $\epsilon_{xc}$ functional (HSE06) with default $\alpha$ = 0.25, i.e., incorporating 25$\%$ of Hartree-Fock exact exchange to capture the electron’s self-interaction error along with SOC. It gives a direct band gap of 85 meV, which is also not in accordance with previously reported theoretical calculations\cite{singh2016role}. Therefore, we have performed G$_0$W$_0$ calculations on top of the orbitals obtained from the PBE+SOC (G$_0$W$_0$@PBE+SOC) and HSE06+SOC (G$_0$W$_0$@HSE06+SOC). The respective band gaps are 280 meV and 249 meV (see Table \ref{T1}). G$_0$W$_0$@PBE+SOC gives the most accurate band gap, however, the band profile is not much affected by the choice of $\epsilon_{xc}$ functional (see Section I of Supplementary Information (SI)). Therefore, we have used PBE $\epsilon_{xc}$ functional to plot the band structures in view of its low computational cost. The band structure of TlBiSe$_2$ with the projected wavefunctions to atomic orbitals is shown in Figure \ref{pic2}(a). The Conduction Band (CB) is dominated by Bi-\textit{p} and Tl-\textit{p} orbitals, while Se-\textit{p} orbitals dominate the Valence Band (VB). The inclusion of SOC has led to an increase in the band gap around the $\Gamma$ point. The valence and conduction band edges switch their orbital character around this point, indicating the band inversion. The \textit{p} orbitals of Se and Bi are involved in this band inversion, as can be clearly seen from Figure \ref{pic2}(b).

\begin{table}[htbp]
	\caption {Band gap (in meV) of TlBiSe$_2$ using different $\epsilon_{xc}$ functionals.}
	\begin{center}
		\begin{adjustbox}{width=0.8\textwidth} 
			\setlength\extrarowheight{+4pt}
			\begin{tabular}[c]{|c|c|c|c|c|} \hline		
				 \textbf{PBE+SOC} & \textbf{HSE06+SOC} & \textbf{G$_0$W$_0$@PBE+SOC}   & \textbf{G$_0$W$_0$@HSE06+SOC} \\ \hline
				  215  & 85& 280 & 249        \\ \hline
			\end{tabular}
		\end{adjustbox}
		\label{T1}
	\end{center}
\end{table}

The similar band inversion has also been observed in TlBiTe$_2$ and TlSbSe$_2$\cite{singh2016role}. 
Following the trend, we have carried out complete substitution in TlBiSe$_2$ at Tl, Bi or Se sites and Tl, Bi sites simultaneously to obtain different materials belonging to the same class (see Figure \ref{pic1}(c)). The band structure of these materials are plotted to see the effect of SOC on the orbital contribution projected on the bands lying near Fermi level.
Firstly, Ga is substituted at the Tl site (Ga$_{Tl}$) to form GaBiSe$_2$. It also crystallizes in \textit{R}$\bar{3}$\textit{m} phase with lattice parameters given as \textit{a} = \textit{b} = \textit{c} = 7.18 \AA \: (in rhombohedral setting) and yields the band structure shown in Figure \ref{pic2}(c). SOC driven inversion of energy levels along with opening of band gap takes place at high symmetry point $\Gamma$ (Figure \ref{pic2}(d)). The projected wavefunctions to atomic orbitals show that \textit{p} orbitals of all atoms contribute near the Fermi level. The inversion involves Bi-\textit{p} and Se-\textit{p} orbitals (similar to the case of TlBiSe$_2$), giving an initial indication that the material can harbor non-trivial topological phase. The indirect band gap as calculated by G$_0$W$_0$@PBE+SOC is 183 meV.

\begin{table}[htbp]
	\caption {Band gap (in meV) of different materials using G$_0$W$_0$@PBE+SOC.}
	\begin{center}
		\begin{adjustbox}{width=0.3\textwidth} 
			\setlength\extrarowheight{+4pt}
			\begin{tabular}[c]{|c|c|} \hline		
				\textbf{Material}    & \textbf{Indirect Band Gap} \\ \hline 
				\textbf{GaBiSe$_2$}  & 183    \\ \hline
			 \textbf{PbBiSe$_2$} & 3     \\ \hline
			 \textbf{SnBiSe$_2$}  & 20   \\ \hline
			 \textbf{SbBiSe$_2$} & 118   \\ \hline
			  \textbf{Bi$_2$Se$_2$}  & 327  \\  \hline
		    \textbf{TlSnSe$_2$}  &  44    \\  \hline
		    \textbf{PbSbSe$_2$} &  41  \\ \hline			
			\end{tabular}
		\end{adjustbox}
		\label{T2}
	\end{center}
\end{table}


After GaBiSe$_2$, we have substituted Pb$_{Tl}$ to get PbBiSe$_2$ having lattice parameters as \textit{a} = \textit{b} = \textit{c} = 8.27 \AA, and crystallizing in the \textit{R}$\bar{3}$\textit{m} phase. G$_0$W$_0$@PBE+SOC yields an indirect band gap of 3 meV. The VB and CB are mainly composed of \textit{p} orbitals of Pb, Bi and Se, as shown in Figure \ref{pic2}(e) and \ref{pic2}(f). The parity inversion occurs at $\Gamma$ and F points, unlike the former. In this BZ, there are 8 TRIMs, i.e., $\Gamma$, Z (non-degenerate) and F, L (triply-degenerate). Therefore, the inversion is occurring at even number of TRIMs, which means that the system should be in trivial state. However, it has been found that if even number of band inversions occur in the first quadrant, but if one or more BZ sides possess odd number of band inversions, then a WTI can be obtained\cite{das2016pedagogic}.

\begin{figure*}[htp]
	\includegraphics[width=0.8\textwidth]{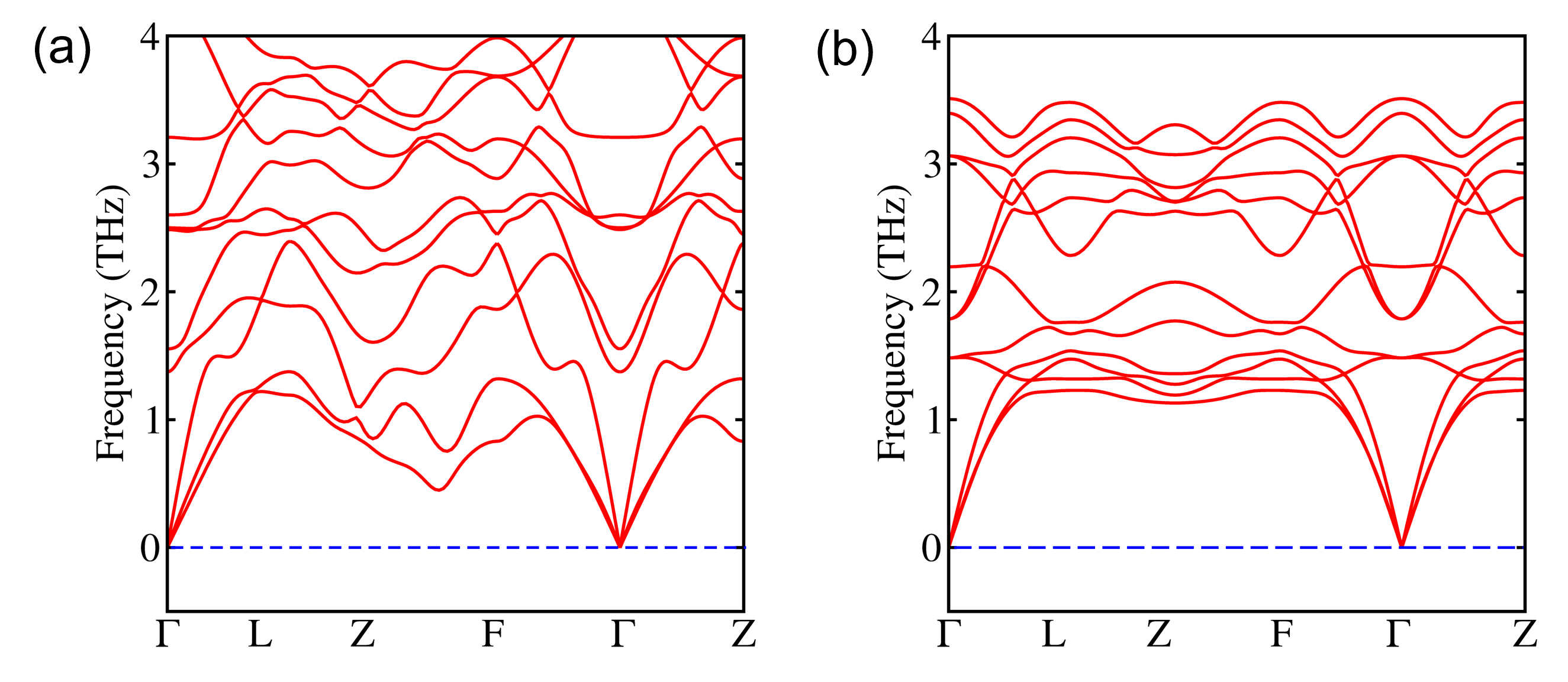}
	\centering
	\caption{Phonon
		band structures of (a) GaBiSe$_2$ and (b) PbBiSe$_2$.}
	\label{pic3}
\end{figure*}

A similar type of parity inversion at even number of TRIMs is obtained for SnBiSe$_2$, SbBiSe$_2$, Bi$_2$Se$_2$, TlSnSe$_2$ and PbSbSe$_2$  (refer Section II and Section III of SI for band structures and optimized lattice parameters, respectively). The band gap of all these materials calculated using G$_0$W$_0$@PBE+SOC are given in Table \ref{T2}. After screening the systems for SOC-induced inversion in the band structures, we have analyzed them for dynamical stability. Figure \ref{pic3} shows the phonon band structures for GaBiSe$_2$ and PbBiSe$_2$. The absence of negative frequencies confirms the dynamical stability of these materials. For other systems viz. SnBiSe$_2$, SbBiSe$_2$, Bi$_2$Se$_2$, TlSnSe$_2$ and PbSbSe$_2$, see Section IV of SI. 




The presence of band inversion on inclusion of SOC is a telltale signature of possibility of non-trivial phase. However, an inverted band structure cannot be considered as a sole criterion to assure the existence of a non-trivial phase. Therefore, further analysis is required to establish its topological nature.
Previous studies have reported that non-trivial band topology generates metallic surface states which are the hallmark of TIs\cite{hsieh2008topological}. In view of this, we have computed the spectrum of surface states by considering a semi infinite slab of 3D material. These lattice surfaces possess Dirac cones, lying at the same \textit{k}-point where the band inversion has occurred in the corresponding bulk band structure. Furthermore, to elucidate the topological nature of the materials, Z$_2$ topological invariants are calculated. TRS yields four distinct Z$_2$ invariants ($\nu_0$;$\nu_1$$\nu_2$$\nu_3$) in 3D case. Each of these four invariants takes up value either 0 or 1, indicating a total of 16 phases with three general classes: a normal insulator, a STI and a WTI\cite{noguchi2019weak}. An ordinary or trivial insulator is obtained when all four invariants are zero, i.e., (0;000), while a STI is obtained when $\nu_0$ = 1. This type of system is robust against weak time-reversal invariant perturbations. However, when $\nu_0$ = 0, and at least one of the indices out of $\nu_1$, $\nu_2$ or $\nu_3$ is nonzero, then the material is WTI. It can be viewed as a stacking of 2D TIs, and is less robust against perturbations.

\begin{table}[htbp]
	\caption {Calculated Z$_2$ invariants for different materials.}
	\begin{center}
		\begin{adjustbox}{width=0.6\textwidth} 
			\setlength\extrarowheight{+4pt}
			\begin{tabular}[c]{|c|c|c|c|} \hline		
				\textbf{Material}  & \textbf{Z$_2$:($\nu_0$;$\nu_1$$\nu_2$$\nu_3$)} & \textbf{Number of surface states} & \textbf{Type}  \\ \hline
				\textbf{GaBiSe$_2$}        &  (1;000)      & 1 & STI       \\ \hline
				\textbf{PbBiSe$_2$}        &  (0;001)      & 2 & WTI        \\ \hline
				\textbf{SnBiSe$_2$}        &  (0;001)      & 2 & WTI        \\ \hline
				\textbf{SbBiSe$_2$}        &  (0;001)      & 2 & WTI        \\ \hline
				\textbf{Bi$_2$Se$_2$}      &  (0;001)      & 2 & WTI         \\ \hline
				\textbf{TlSnSe$_2$}        &  (0;001)      & 2 & WTI         \\ \hline
				\textbf{PbSbSe$_2$}        &  (0;111)      & 2 & WTI         \\ \hline
			\end{tabular}
		\end{adjustbox}
		\label{T3}
	\end{center}
\end{table}

\begin{figure*}[htp]
	\includegraphics[width=0.8\textwidth]{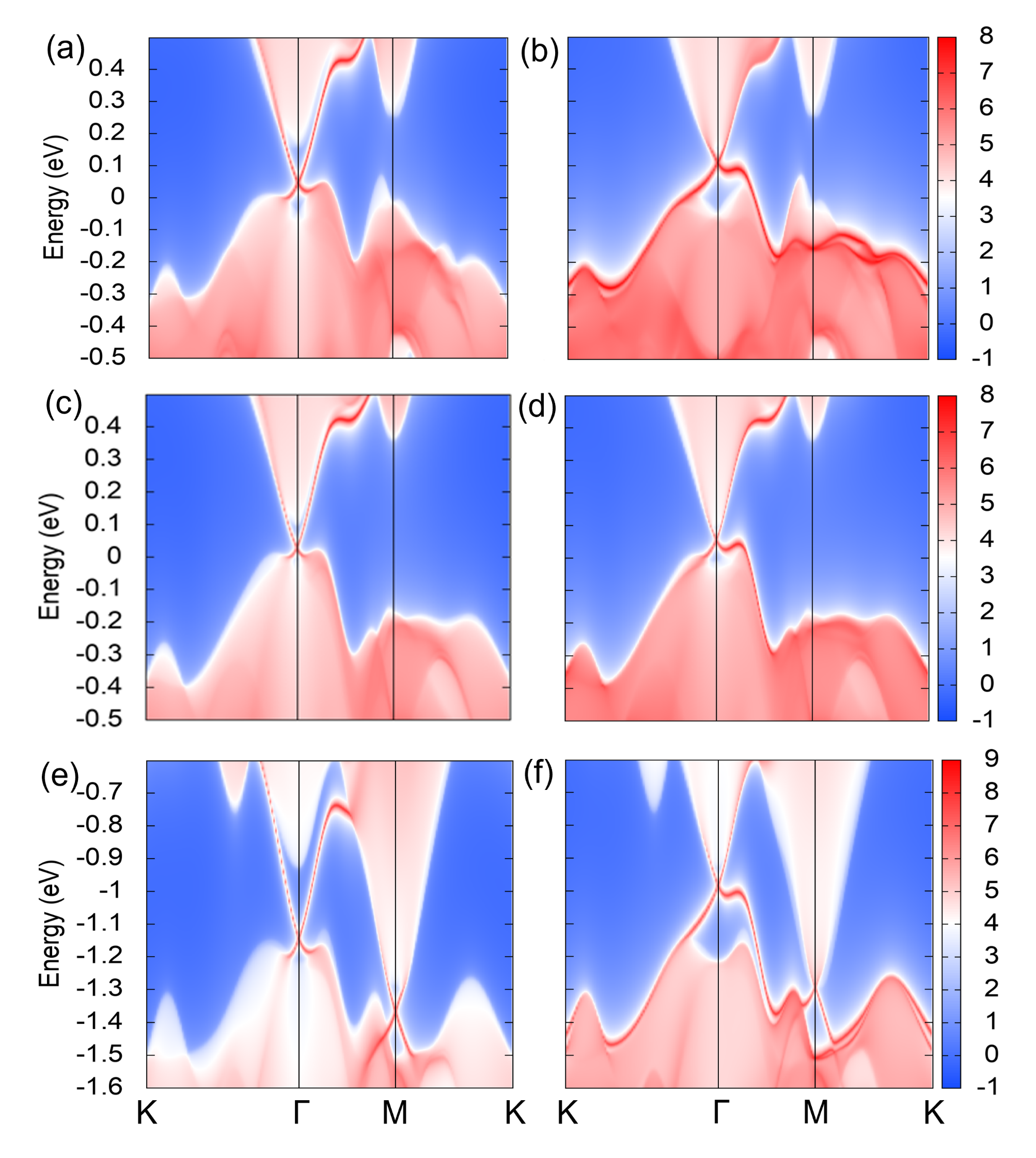}
	\centering
	\caption{Surface states for the left surface of TlBiSe$_2$, GaBiSe$_2$ and PbBiSe$_2$ are shown in (a), (c), (e) and for the right surface are shown in (b), (d), (f), respectively. Here, the sharp red curves represent surface states, whereas, the shaded regions show the spectral weight of projected bulk bands. }
	\label{pic4}
\end{figure*}

It has already been established that TlBiSe$_2$ is a strong 3D TI\cite{kuroda2010experimental}, on that account, we first obtain its surface band structure. For this, a tight-binding Hamiltonian with MLWFs considering the projection of \textit{p} orbitals of Bi/Se and \textit{sp} orbitals of Tl is constructed. Since the left and right surface for TlBiSe$_2$ terminates with different atoms (Tl and Se, respectively), therefore, we have plotted surface state spectra of (111) surface for both the surface terminations. Figure \ref{pic4}(a) and \ref{pic4}(b) show a single surface state protected by TRS at the $\Gamma$ point in the projected 2D BZ. Alongside, we have calculated the topological invariant, $\nu_0$, which comes out to be 1, confirming that TlBiSe$_2$ is a STI. Following this, we have explored GaBiSe$_2$ for topological properties. In this case, the \textit{p} orbitals of Bi/Se and \textit{sp} orbitals of Ga are considered in constructing the tight-binding Hamiltonian. A single Dirac cone protected by TRS has been observed for (111) surface (see Figure \ref{pic4}(c) and \ref{pic4}(d)), and the corresponding topological invariant is (1;000), which is a signature of non-trivial topology. Hence, to the best of our knowledge, we report here GaBiSe$_2$ to be a STI presumably for the first time.

Afterwards, we have performed surface band structure calculation for PbBiSe$_2$. We have obtained even (two) number of surface states lying at $\Gamma$ and M (see Figure \ref{pic4}(e) and \ref{pic4}(f)). The presence of even number of surface states yields $\nu_0$ = 0, and the weak indices come out to be $\nu_1$ = 0, $\nu_2$ = 0 and $\nu_3$ = 1. As even number of surface states lead to scattering and thus are not topologically protected, hence this material is categorized as a WTI. The similar calculations are performed for SnBiSe$_2$, SbBiSe$_2$, Bi$_2$Se$_2$, TlSnSe$_2$ and PbSbSe$_2$. All of them show even number of surface states, as shown in Section V of SI, and corresponding Z$_2$ invariants are given in Table \ref{T3}. These materials belong to the class of Z$_2$ WTI. Nowadays, WTIs are also getting attention as it has been found that their surface states are robust against imperfections, owing to the delocalization of surface electrons\cite{ringel2012strong}. 


In summary, we have performed isostructural substitutions of materials based on SOC-induced parity inversion in the band structure. Using first-principles based methodologies, viz., PBE, HSE06, and many-body perturbation theory (G$_0$W$_0$), we have systematically studied the electronic structure of topological materials belonging to \textit{R}$\bar{3}$\textit{m} space group. The band gap calculated using G$_0$W$_0$@PBE+SOC is in close agreement with the experimental value. We have revealed that GaBiSe$_2$ is a STI as its surface accommodates a single crossing at the $\Gamma$ point. PbBiSe$_2$, SnBiSe$_2$, SbBiSe$_2$, Bi$_2$Se$_2$, TlSnSe$_2$ and PbSbSe$_2$ are WTI catering even number of surface states within the bulk band gap. The absence of negative frequencies in the phonon band structures indicates dynamical stability. The calculated Z$_2$ invariants are in accordance with the surface state plots confirming their topological nature. Discovery of these materials will offer a great platform for studying novel quantum effects.

\section{Compulational Methods}
The calculations are performed using DFT\cite{hohenberg1964inhomogeneous,kohn1965self} with the Projected Augmented Wave (PAW)\cite{kresse1999ultrasoft,blochl1994projector} method implemented in Vienna \textit{Ab initio} Simulation Package (VASP)\cite{kresse1996efficient} code. All the structures are optimized with the Generalized Gradient Approximation (GGA) of Perdew-Burke-Ernzerhof (PBE)\cite{perdew1996generalized} until the Hellmann-Feynman forces are smaller than 0.001 eV/\AA. The plane wave basis is used with 400 eV cutoff energy. The $\Gamma$-centered 6$\times$6$\times$4  \textit{k}-grid is used to sample the irreducible BZ of rhombohedral phase with the \textit{R}$\bar{3}$\textit{m} space group. SOC is included in all calculations except in ionic optimization. The advanced hybrid $\epsilon_{xc}$ functional (HSE06) including SOC and many-body perturbation methods,  G$_0$W$_0$@PBE+SOC and G$_0$W$_0$@HSE06+SOC are used for the better estimation of the band gap\cite{hedin1965new}. For this, we have used  4$\times$4$\times$4 \textit{k}-grid, and the number of bands is set to six times the number of occupied bands. The phonon calculations are performed with 4$\times$4$\times$4 supercells using the PHONOPY package\cite{togo2008first,togo2015first}. In order to investigate the topological properties of the materials, we have performed DFT calculations using fully relativistic norm-conserving pseudopotentials as implemented in the QUANTUM ESPRESSO code\cite{giannozzi2009quantum}. The results of these DFT calculations are then fed as input to WANNIER90\cite{mostofi2008wannier90} for constructing a tight-binding model based on Maximally Localized Wannier Functions (MLWFs) with \textit{p} orbitals of Se, Bi and \textit{sp} orbitals of Ga, Tl, Sn, Sb, Pb. The surface states and topological invariants are then calculated using the  Green’s function method as implemented in the Wannier-TOOLS package\cite{wu2018wanniertools}.

	


\begin{acknowledgement}
	A.P. acknowledges IIT Delhi for the junior research fellowship. P.B. acknowledges UGC, India, for the senior research fellowship [Grant no. 1392/(CSIR-UGC NET JUNE 2018)]. M.J. acknowledges CSIR, India, for the senior research fellowship [Grant No. 09/086(1344)/2018-EMR-I]. S.B. acknowledges financial support from SERB under a core research grant (grant no. CRG/2019/000647) to set up his High Performance Computing (HPC) facility ‘‘Veena’’ at IIT Delhi for computational resources.
\end{acknowledgement}
\begin{suppinfo}
Band profile of TlBiSe$_2$ using PBE+SOC and G$_0$W$_0$@PBE+SOC $\epsilon_{xc}$ functionals; Band structures of SnBiSe$_2$, SbBiSe$_2$, Bi$_2$Se$_2$, TlSnSe$_2$ and PbSbSe$_2$; Optimized lattice parameters of  SnBiSe$_2$, SbBiSe$_2$, Bi$_2$Se$_2$, TlSnSe$_2$ and PbSbSe$_2$; Phonon band structures of SbBiSe$_2$, SnBiSe$_2$, Bi$_2$Se$_2$, TlSnSe$_2$ and PbSbSe$_2$; Surface state band structures of SnBiSe$_2$, SbBiSe$_2$, Bi$_2$Se$_2$, TlSnSe$_2$ and PbSbSe$_2$; Band gap (in meV) of different materials using G$_0$W$_0$@HSE06+SOC $\epsilon_{xc}$ functional.
	
\end{suppinfo}
\providecommand{\latin}[1]{#1}
\makeatletter
\providecommand{\doi}
{\begingroup\let\do\@makeother\dospecials
	\catcode`\{=1 \catcode`\}=2 \doi@aux}
\providecommand{\doi@aux}[1]{\endgroup\texttt{#1}}
\makeatother
\providecommand*\mcitethebibliography{\thebibliography}
\csname @ifundefined\endcsname{endmcitethebibliography}
{\let\endmcitethebibliography\endthebibliography}{}

\end{document}